\documentclass[prb,twocolumn,superscriptaddress,floatfix,nopacs]{revtex4}
\usepackage{amsfonts,amsmath,epsf,dcolumn,natbib}
\usepackage{color}

\newcommand{\half}{\tfrac{1}{2}}

\newcommand{\be}{\begin{equation}}
\newcommand{\ee}{\end{equation}}
\newcommand{\bea}{\begin{eqnarray}}
\newcommand{\eea}{\end{eqnarray}}
\newcommand{\balg}{\begin{align}}
\newcommand{\ealg}{\end{align}}

\begin{document}

\title{Trivial Constraints on Orbital-free Kinetic Energy Density Functionals}
\author{Kai Luo}
\email{kluo@ufl.edu}
\affiliation{Quantum Theory Project, Department of Physics, University of Florida, Gainesville, FL 32611}
\author{S.B.~Trickey}
\email{trickey@qtp.ufl.edu}
\affiliation{Quantum Theory Project, Departments of Physics and
of Chemistry, University of Florida, Gainesville, FL 32611}

\date{13 Dec.\ 2017; revised 17 Jan.\ 2018}

\begin{abstract}
Approximate kinetic energy density functionals (KEDFs) are central to
orbital-free density functional theory. Limitations on the spatial
derivative dependencies of KEDFs have been claimed from differential
virial theorems.  We identify a central defect in the argument: the
relationships are not true for an arbitrary density but hold only for
the minimizing density and corresponding chemical potential. Contrary
to the claims therefore, the relationships are not constraints and
provide no independent information about the spatial derivative
dependencies of approximate KEDFs.  A simple argument also shows that
validity for arbitrary $v$-representable densities is not restored by
appeal to the density-potential bijection.


\end{abstract}
                        
\maketitle

\section{\label{Intro}Introduction}

Unarguably the dominant contemporary form of many-electron theory for
computing the properties of complicated molecules, clusters, and
almost all extended systems is density functional theory (DFT) in its
Kohn-Sham (KS) form \cite{KS}.  Conventionally the KS scheme is used
to render the DFT Euler equation in the form of a mean-field orbital
eigenvalue problem, the KS equations.  Though enormously successful,
this approach has the standard computational cost barrier of any
eigenvalue problem, namely cubic cost scaling with the number of
electrons (or equivalent, the number of basis functions).  That
motivates long-standing interest in orbital-free DFT (OF-DFT)
\cite{VVKDCSBTreview,WesolowskiWangBook}, which in principle scales
with system size.

OF-DFT, however, introduces the challenge of approximating
the KS kinetic energy (KE) as an explicit density functional, e.g., 
\be
T_{\mathrm s}[n] := \int d{\mathbf r}\, t_{\mathrm s}[n({\mathbf r})] \; ,
\label{1ptfnal}
\vspace*{-3pt}
\ee
instead of the familiar orbital-dependent version
\begin{eqnarray}
T_{\rm s}[\{\varphi_i\}_{i=1}^{N_e}]
&:=& \half \sum_{i=1}^{N_e}\int\, d{\mathbf r} \, 
|\nabla \varphi_i({\mathbf r})|^2 \nonumber\\
&\equiv& \int d{\mathbf r} \, t^{\mathrm{orb}}_{\rm s}({\mathbf r}) 
\label{A1}
\end{eqnarray}
in Hartree atomic units. 
[Remark: In this form the integrand is manifestly positive definite. The 
more common Laplacian form is not. The difference is a
surface integral which ordinarily is zero.] Here $N_e$ is the number of
electrons and the ground state number density is %
\be
n_0({\mathbf r}) = \sum_i^{N_e} f_i\, | \varphi_i ({\mathbf r})|^2 \, .
\label{densitydefn}
\ee
where the spin-orbital occupation numbers, $f_i$, at zero temperature are $0$ or $1$, except for the case of degeneracy at the Fermi level\cite{DreizlerGrossBook}.  

Orbital-free DFT  aims to provide useful approximations 
to $T_{\mathrm s}[n]$  without explicit use of the KS orbitals.  If one restricts 
attention to single-point approximations, $t_{\mathrm s}^{approx}[n({\mathbf r}]$,
a basic issue is the maximum order of spatial derivative
dependence to be included. Generalized 
gradient approximations \cite{PRB88} (GGA) and Laplacian-level 
functionals \cite{PerdewConstantin07,LarrichiaEtAl2014,CancioStewartKuna2016,CancioRedd2017} are the practical limits so far.  Various  
dimensionless spatial derivative combinations (reduced density derivatives) 
have been proposed \cite{PRB80} but little is known about how
to select from among them. An exception would seem to be 
papers by Baltin \cite{Baltin1987} and co-workers \cite{ShaoBaltin1990}
and others \cite{HolasMarch1995,AlharbiKais2017}.  
Those use differential virial theorems to derive constraints on the 
order of spatial derivative that can appear.  

Here we show that those 
relationships are not constraints but trivial identities of complicated
form satisfied only by the equilibrium density (i.e. ground-state density) 
for a given external
potential $v_{\mathrm{ext}} = \delta E_{\mathrm{ext}}/\delta n$.

We begin the next section with the pertinent aspects of the KS Euler
equation.  Then we rehearse the original arguments from
Ref.\ [\onlinecite{Baltin1987}] using the one-dimensional (1D) case presented 
there. (The three-dimensional case uses identical logic but
is more cumbersome, so we do not treat it explicitly.)  In
the subsequent section, we discuss two related omissions in those
arguments which significantly alter the claimed consequences to the
point of triviality.  We illustrate by reconsidering two cases
originally treated in Ref.\ [\onlinecite{Baltin1987}].  Brief
consideration to show that a seemingly plausible Hohenberg-Kohn
bijectivity argument does not alter the result concludes the
presentation.

\section{\label{DiffVirCon1D}Differential Virial Constraint- 1D}
\vspace*{-3pt} \subsection{Euler Equation}

The KS decomposition of the universal ground-state
total electronic energy density functional is \cite{KS}   
\begin{equation}
E[n]=T_{\rm s}[n]+E_{\rm ext}[n]+E_{\rm H}[n] + E_{\rm xc}[n] \; ,
\label{A2}
\end{equation}
with $T_{\rm s}[n]$ the non-interacting kinetic energy functional as defined
above,   $E_{\rm ext}[n]$ the external field interaction energy,  
$E_{\rm H}[n]$ the Hartree energy (classical
electron-electron repulsion), and $E_{\rm xc}[n]$ the exchange-correlation (XC)
energy functional. (Remark: any external system configurational
energy, e.g., ion-ion repulsion, is omitted as irrelevant here.)  
Minimization gives a single Euler equation, 
\begin{equation}
\frac{\delta T_{\rm s}[n]}{\delta n({\bf r})}+v_{\rm KS}([n];{\bf r})=\mu .
\label{A3}
\end{equation}
Here $v_{\rm KS}=\delta (E_{\rm ext} + E_{\rm H} + E_{\rm xc})/\delta n$ is the KS
potential and $\mu$ is the chemical potential such that the 
minimizing density $n_0$ yields the correct $N_e$.   Explicit use
of the KS KE orbital dependence renders the Euler equation as 
the familiar KS equation 
\be
\lbrace -\half \nabla^2 + v_{\mathrm{KS}}([n];{\mathbf r})\rbrace \varphi_i({\mathbf r}) = \epsilon_i \varphi_i ({\mathbf r}) \, .
\label{ordinaryKS}
\ee

\vspace*{-3pt}\subsection{Original Differential Virial Argument}

The original argument of Ref.\ [\onlinecite{Baltin1987}] 
follows in our notation.   Consider a 1D system
and its KS potential and states.  For it the differential virial theorem 
(Eq.\ (13) of Ref.\ [\onlinecite{Baltin1985}]) is
\be
\frac{d t_{\mathrm s}(x)}{dx} =  \frac{1}{8}\frac{d^3 n(x)}{dx^3} - %
\half n(x) \frac{d v_{\mathrm{KS}}}{dx}  
\label{Baltin1DHVTa}
\ee
or in primed notation as used in Ref.\ [\onlinecite{Baltin1987}],  
\be
t^{\prime}_{\mathrm s}(x) = \frac{n^{\prime\prime\prime}(x)}{8} -\half n(x) v^{\prime}_{\mathrm {KS}}(x) \; .
\label{Baltin1DHVTb}
\ee
(Remark: To get to the Euler equation equivalent to our Eq.\ (\ref{A3}), Eq.\ (7)
of Ref.\ [\onlinecite{Baltin1987}] writes the supposed equivalent of our 
Eq.\ (\ref{A2}) in 1D as
\be
E[n] = \int_{-\infty}^{\infty} dx \, t_{\mathrm s}(x) + \int_{-\infty}^{\infty} dx \, n(x) %
v_{\mathrm {KS}}(x) \; .
\label{Baltin1DEdft}
\ee
This is incorrect since $v_{\mathrm{KS}}$ is not solely the external potential 
but the error is inconsequential for the discussion at hand.)   

Ref.\  [\onlinecite{Baltin1987}] then considers a one-point approximation 
for $t_{\mathrm s}$ that 
depends on spatial derivatives of $n$ through $g^{th}$ order:
\be
t_{\mathrm s}(x) := {\mathsf f}(n, n^\prime, n^{\prime\prime}, n^{\prime\prime\prime}, \ldots 
n^{(g)})  \; . 
\label{gthorderfnal}
\ee
Straightforwardly one gets 
\be
t^{\prime}_{\mathrm s}(x) = \sum_{\nu=0}^{g} \frac{\partial {\mathsf f}}{\partial n^{(\nu)}}\frac{d n^{(\nu)}}{dx} = \sum_{\nu=0}^{g} \frac{\partial {\mathsf f}}{\partial n^{(\nu)}} n^{(\nu+1)}  \; 
\label{BaltinEq11}
\ee
which is Eq.\ (11) in Ref.\ [\onlinecite{Baltin1987}].  Alternatively,
repeated integration by parts gives Eq.\ (6) of that reference, 
\be
\frac{\delta T_{\mathrm s|}}{\delta n} = \sum_{\nu =0}^{g}(-1)^\nu %
\frac{d^\nu}{dx^\nu}\left\lbrack \frac{\partial {\mathsf f}}{\partial n^{(\nu)}} \right\rbrack  \; .  
\label{BaltinEq6}
\ee

Ref.\ [\onlinecite{Baltin1987}] then rewrites the Euler equation 
(\ref{A3}) with (\ref{BaltinEq6}) and takes one spatial derivative to get 
\be
v^\prime_{\mathrm KS}(x) = \sum_{\nu =0}^{g}(-1)^{\nu +1} %
\frac{d^{\nu+1}}{dx^{\nu+1}}\left\lbrack \frac{\delta {\mathsf f}}{\delta n^{(\nu)}} \right\rbrack  \; .
\label{BaltinVprime}
\ee
Substitution of both this result and the result from (\ref{BaltinEq11}) 
in Eq.\ (\ref{Baltin1DHVTb}) then gives 
\be
\sum_{\nu =0}^{g}\left\lbrack(-1)^{\nu} %
    n \frac{d^{\nu+1}}{dx^{\nu+1}}\left( \frac{\partial {\mathsf f}}{\partial n^{(\nu)}} \right) %
- 2 n^{(\nu +1)} \frac{\partial {\mathsf f}}{\partial n^{(\nu)}} \right\rbrack = %
-\frac{1}{4}n^{\prime\prime\prime} \; .
\label{BaltinEq10}
\ee

Ref.\ [\onlinecite{Baltin1987}] then says that ``\ldots this equation 
has to be looked upon as a relation to be satisfied identically with respect
to the variables $n$, $n^\prime$, \ldots $n^{(\ell)}$ occurring in it and that
the equation ``is a condition
to be imposed on the dependence of $\mathsf f$ upon the 
variables $n$, $n^\prime$, \ldots $n^{(\ell)}$''.  There follows 
an examination of 
functions ${\mathsf f}$ which depend on $n^{(g)}$ through $g=2$
with the conclusion that the only allowable form consists of the
full von Weizs\"acker term \cite{Weizsacker} plus an arbitrarily 
scaled Thomas-Fermi term \cite{Thomas,Fermi}. 

\subsection{Difficulty}

There are two consequential flaws in the foregoing argument that seem
not to have been recognized heretofore.  They have a common stem.
 First, the differential virial relation from which
Eq.\ (\ref{Baltin1DHVTb}) is derived holds {\it only} for the exact
eigenstates of the given Hamiltonian.  In the KS case with fixed
external potential, that differential virial relation therefore properly
reads
\be
t^{\prime}_{\mathrm s}(n_0(x)) = \frac{n_0^{\prime\prime\prime}(x)}{8} -\half n_0(x) v^{\prime}_{\mathrm KS}(n_0(x)) \; .
\label{Baltin1DHVTbproper}
\ee

The same error occurs in use of the Euler equation to get the 
spatial derivative of the potential.  The Euler equation is not
a general functional relation for arbitrary density $n$.  Rather it
is a relationship between the minimizing density $n_0$ and the 
unique (up to a constant) external potential which is paired with 
that $n_0$.  Thus Eq.\ (\ref{BaltinVprime}) must be replaced by
\be
v^\prime_{\mathrm KS}(n_0(x)) = \sum_{\nu =0}^{g}(-1)^{\nu +1} %
\frac{d^{\nu+1}}{dx^{\nu+1}}\left\lbrack \frac{\delta {\mathsf f}}{\delta n^{(\nu)}} \right\rbrack_{n_0}  \; .
\label{BaltinVprimeproper}
\ee
As a consequence, the purported constraint on functional dependence
becomes 
\bea
\sum_{\nu =0}^{g} && \left\lbrack(-1)^{\nu} %
    n_0\, \frac{d^{\nu+1}}{dx^{\nu+1}}\left( \frac{\partial {\mathsf f}}{\partial n^{(\nu)}} \right)_{n_0} \right. \nonumber \\
&& \left. - 2 n_0^{(\nu +1)} \frac{\partial {\mathsf f}}{\partial n^{(\nu)}}\Big\vert_{n_0} \right\rbrack = %
-\frac{1}{4}n_0^{\prime\prime\prime} \; .
\label{BaltinEq10proper}
\eea
This is a requirement on the behavior of $\mathsf f$ at a single point $n_0$ 
in the space of one-body densities and paired with a specific $v_{\mathrm {ext}}$.   
Contrary to Ref.\ [\onlinecite{Baltin1987}],
Eq.\ (\ref{BaltinEq10proper})  is {\it not} a condition on the  
dependence of $\mathsf f$ upon the 
variables $n$, $n^\prime$, \ldots $n^{(\ell)}$ for arbitrary density $n$ 
given a $v_{\mathrm {ext}}$.  Rather, given a dependence through order $n^{(\ell)}$,
and a particular $v_{\mathrm{ext}}$, the requirement is to find the 
equilibrium density 
$n_0$ that satisfies (\ref{BaltinEq10proper}).  
  
\section{\label{Examples1D}1D Examples}

Just as with the original argument, early examples
of the implications of the purported constraint were for 
1D systems.  We analyze two of those early 1D cases as particularly 
clear instances of the trivial nature of the supposed constraint.

\subsection{1D Homogeneous Electron Gas}

For the 1D homogeneous electron gas (HEG), the Thomas-Fermi functional, $T_{TF}$ 
\be
T_{TF}=c_{TF}\int dx\, n^{3}(x)
\label{Ttf1d}
\ee
is exact. Secure in that knowledge, one can put it aside for a moment and
simply consider $T_{TF}$ as a candidate approximate KE functional.  
The associated kinetic energy density and partial 
derivative are 
\bea
t_{TF}&=&c_{TF}n^{3}:={\mathsf f}(n)  \label{TFked} \\
\frac{\partial {\mathsf f}}{\partial n}&=&3c_{TF}n^{2} \; .
\eea
Then Eq.\ (\ref{BaltinEq10proper}) becomes 
\be
n_0\ \frac{d}{dx}(3c_{TF}n_0^{2})-2n_0^{(1)}(3c_{TF}n_0^{2})=-\frac{1}{4}n_0^{(3)}.
\ee
Its solution is
\be
n_0(x)=\frac{1}{2}ax^{2}+bx+c \; ,
\ee
with coefficients to be determined. An appropriate boundary 
condition is periodic
\be
n_0(x+L)=n_0(x) 
\ee
where $L$ is a suitable length.  As a result 
\be
a=b=0 \; .
\label{n0pbc}
\ee
The constant $c$ is set by imposition of
the desired value of the uniform density.  The outcome of the 
supposed constraint is simply to 
demonstrate that $t_{TF}$ is compatible with the HEG.  

If, on the other hand, one 
imposes box boundary (BB) conditions
\be
n_{0}(0) = n_0(L)=0 
\ee
one has $c=0$ and 
\be
n_{0,BB}(x) = \frac{6N_e}{L^3} x (L-x)  \; .
\label{n0box}
\ee
This density, however is unacceptable, since it violates Lieb's condition
\cite{LiebIJQC1984} for the finitude of the KE:
\bea
\int_0^L dx \left(\frac{d \phi}{dx}\right)^2 &<& \infty \label{LiebCondition} \\
 \phi(x) &:=& \sqrt{n_{0,BB}(x)} \label{phidefn}
\eea
Alternatively, one may see the problem with $n_{0,BB}$ by attempting 
direct inversion of the Schr\"odinger equation for $\phi$ in the $N_e=1$ 
case to recover the one-body potential.  Up to a constant, the purported 
potential is negative definite 
with poles at $x=0$, $L$: ($-L^2/[8 x^2(L-x)^2]$), i.e.\ , 
$n_{0,BB}$ is not $v-$representable. There is nothing special about
$N_e=1$ to rescue the case.  

Thus, all that Eq.\ (\ref{BaltinEq10proper}) yields in the $T_{TF}$
case is confirmation that $T_{TF}$ is indeed correct for the 1D HEG.
One also learns that Eq.\ (\ref{BaltinEq10proper}) has
solutions which upon detailed inspection do not correspond to any
potential, but that says nothing about limits on the validity of $T_{TF}$ 
as an approximate functional.  For cases in which $v_{\mathrm{ext}}$ does exist,
Eq.\ (\ref{BaltinEq10proper}) has no information about it and provides
no information on the accuracy of the approximation $t_s \approx
t_{TF}$.  Thus, contrary to Ref.\ [\onlinecite{Baltin1987}], no
general requirement on the dependence of $T_s[n]$ upon spatial
derivative order is obtained from Eq.\ (\ref{BaltinEq10proper}) when
$T_{TF}$ is put to the test.

\subsection{One Electron in 1D}
For a 1-electron system, the von Weizs\"acker functional $T_{W}$
\be
T_{\mathrm s}= T_W = \int\,dx\, t_W(x) 
\label{vWKE}
\ee
is exact.  Its kinetic energy density is
\be
t_{W}=\frac{[n^\prime(x)]^{2}}{8n(x)}:={\mathsf f}_W(n,n^\prime).
\label{tWKED}
\ee
For convenience, the relevant partial derivatives for use of
Eq.\ (\ref{BaltinEq10proper}) are
\begin{eqnarray*}
\frac{\partial {\mathsf f}_W}{\partial n} & = & -\frac{(n^\prime)^{2}}{8n^{2}}\\
\frac{\partial {\mathsf f}_W}{\partial n^\prime} & = & \frac{n^\prime}{4n}
\label{fWderivs}
\end{eqnarray*}
Substitution of these results in the left-hand side of Eq.\ (\ref{BaltinEq10proper}) gives 
\begin{eqnarray*}
\Big(n_0\frac{d}{dx}\frac{-(n_0^\prime)^{2}}{8n_0^{2}}-2n_0^\prime\frac{-(n_0^\prime)^{2}}{8n_0^{2}}\Big) \nonumber \\ 
+\Big(-n_0\frac{d^{2}}{d x^{2}}\frac{n_0^\prime}{4n_0}-2n_0^{\prime\prime}\frac{n_0^\prime}{4n_0}\Big) & = & \\  
\Big(-\frac{n_0^\prime n_0^{\prime\prime}}{4n_0}+\frac{(n_0^\prime)^{3}}{4n_0^{2}}+\frac{(n_0')^{3}}{4n_0^{2}}\Big)\nonumber \\
+\Big(-\frac{1}{4}n_0^{(3)}+\frac{3n_0^\prime n_0^{\prime\prime}}{4n_0}-\frac{(n_0^{\prime})^{3}}{2n_0^{2}}-\frac{n_0^{\prime}n_0^{\prime\prime}}{2n_0}\Big) & = & \\
 -\frac{1}{4}n_0^{(3)}\\
\label{Baltin10forTw}
\end{eqnarray*}
This is the same as the right hand side of Eq.\ (\ref{BaltinEq10proper})
so that equation reduces to a trivial identity for all equilibrium densities 
associated with the combination $T_W$, some $E_{\mathrm{xc}}$, and 
some $E_{\mathrm{ext}}$.  Therefore, no information 
is provided by the differential virial constraint, 
Eq.\  (\ref{BaltinEq10proper}), about 
the functional dependence of  
$T_s$ upon spatial derivatives except that $t_W$ is a valid form.
Note also that unlike the 1D HEG or box-bounded $t_{TF}$ cases considered 
above, there are infinitely many densities that lead to the
trivial identity because there are 
infinitely many single-electron external potentials. Thus, there is no 
access to a unique solution $n_0$ provided by the purported constraint.  \\

\section{\label{Concl}Discussion and Conclusions}

Examination of the 3D version of the differential virial 
constraint argument as summarized,
for example, in Ref.\ [\onlinecite{AlharbiKais2017}], shows that the
same critical mis-use of the Euler equation occurs in 3D as in 1D.  
The preceding analysis therefore holds unchanged.  

It  might seem that Hohenberg-Kohn bijectivity 
between $v_{\mathrm {ext}}$ and $n_0$ could rescue the 
argument by making 
the Euler equation true for an arbitrary v-representable density. Note that
even if that were the case, the differential virial theorem part of
the argument itself 
still would hold only for the extremalizing density.  
But HK bijectivity does not remove 
the Euler equation restriction either.  Bijectivity   
is true for an arbitrary density 
precisely and only in the case that the arbitrarily chosen
density is paired with the unique external potential for which 
it is the {\it minimizing} density 
$n_0$. Bijectivity is irrelevant for the case of interest, 
namely a {\it fixed} $v_{\mathrm{ext}}$ 
and {\it arbitrary} $n({\mathbf r})$. The required pairing of density
and potential is missing.  Thus, though the Euler equation 
holds for arbitrary $n_0$ with the associated $v_{\mathrm {ext}}[n_0]$
and corresponding $\mu[n_0]$, the flaw identified above persists.  
Spatial differentiation of the Euler equation to replace $v^\prime_{\mathrm{KS}}$
(recall Eq.\ (\ref{BaltinVprime})) in the differential virial relation still
ties the result to equilibrium densities $n_0$, not arbitrary
ones.  

It is worth noting that the non-uniqueness of
KE densities \cite{SimLarkinBurkeBock2003} (indeed, any energy density) 
should raise suspicions about the validity of any supposed constraint
on the spatial derivative dependence of an approximation for $t_s$.  
The well-known vanishing of $\nabla^2 n$ terms is an example.   In fact,
it is the counterexample to the Ref.\ [\onlinecite{Baltin1987}] 
argument (just after Eq.\ (48) of that reference) that the approximate 
KE density must obey $t_s^{approx} \ge 0$.  That constraint is highly
valuable but it is a choice of gauge.  The issue is discussed in
detail in Refs.\ [\onlinecite{APN2002,TKJ2009}].  A related 
issues is that it is not self-evident that a function having
up through g$^{th}$ order derivatives, Eq.\ (\ref{BaltinEq11}), necessarily
is itself differentiable for arbitrary densities. Nor is it always
true that one can do the repeated integration by parts, Eq.\ (\ref{BaltinEq6})
with vanishing surface terms; see Ref.\ [\onlinecite{PerdewSahniHarbolaPathak1986}] for counterexamples.

The present analysis resolves at least one other peculiar finding 
in Refs.\ [\onlinecite{Baltin1987}] and [\onlinecite{ShaoBaltin1990}].
Those claim to show that through second-order spatial derivatives
the only KE density form consistent with the supposed differential virial 
constraint is  $t_s^{approx} = \lambda t_{TF} + t_{W}$ with $\lambda$ 
an undetermined constant. This restriction is suspect on its
face because of the Lieb conjecture  
\cite{Lieb80} that $T_s \le T_{TF}+T_{W}$.  That conjecture is consistent 
with the   $N_e \rightarrow \infty$ limit of the bound found by 
G\'azquez and 
Robles \cite{GazquezRobles82}. (See also Acharya et al. \cite{AcharyaEtAl1980} for a heuristic
formulation with number dependence that has the Lieb bound as the $N_e \rightarrow \infty$ limit.) For finite systems, the  
only straightforward way to make the peculiar result consistent with
the G\'azquez-Robles expression would be for $\lambda$ to be number-dependent,
thereby raising an obvious problem of size-consistency. 
There is no obvious simple way to make the result 
consistent with the Lieb bound in the thermodynamic limit.  The analysis
presented here removes that problem by showing that all the 
supposed constraint 
really does is to confirm that for a specific $n_0$ one always can find 
a $\lambda$ which makes the claim true.  Just pick    
\be
\lambda[n_0] = \frac{T_{\mathrm s}[n_0] - T_W[n_0]}{T_{TF}[n_0]}  \; .  
\ee
While true, it is essentially tautological, hence useless. \\

\begin{acknowledgments}
We acknowledge, with thanks, informative conversations with Jim Dufty and 
Valentin Karasiev and a helpful email exchange with Paul Ayers.  
This work was supported  by the U.S.\ Dept.\ of Energy 
grant DE-SC0002139.  
 
\end{acknowledgments}

\end{document}